\documentclass[twocolumn,aps,prc,superscriptaddress,showpacs,floatfix]{revtex4}

\usepackage{epsfig}
\usepackage{amsmath}
\usepackage{multirow}
\usepackage{graphicx}
\usepackage{amsmath}
\usepackage{feynmf}
\usepackage[normalem]{ulem}  
\usepackage[dvips]{color} 

\renewcommand\sout{\bgroup \color{red} \ULdepth=-.5ex \ULset}

\newcommand{\Ex}[2]{\ifmmode{#1\times10^{#2}}\else{$#1\times10^{#2}$}\fi}

\DeclareMathAlphabet{\mathpzc}{OT1}{pzc}{m}{it}

\newcommand{\bra}{\langle}
\newcommand{\ket}{\rangle}
\newcommand{\fig}[1]{Fig.\,\ref{#1}}

\newcommand{\eqn}[1]{Eq.\,(\ref{#1})}

\newcommand{\ud}{\mathrm{d}}

\begin{document}


\title{A renewed look at $\eta^\prime$ in medium}

\author{Youngshin Kwon}
\email{ykwon@ssu.ac.kr}
\affiliation{Department of Physics, Soongsil University, Seoul 156-743, Korea}
\affiliation{Department of Physics and Institute of Physcis \& Applied Physics, Yonsei University, Seoul 120-749, Korea}

\author{Su Houng Lee}
\email{suhoung@yonsei.ac.kr}
\affiliation{Department of Physics and Institute of Physcis \& Applied Physics, Yonsei University, Seoul 120-749, Korea}

\author{Kenji Morita}
\email{kmorita@yukawa.kyoto-u.ac.jp}
\affiliation{Yukawa Institute for Theoretical Physics, Kyoto University, Kyoto 606-8502, Japan}

\author{Gy\"orgy Wolf}
\email{wolf.gyorgy@wigner.mta.hu}
\affiliation{Wigner RCP, RMKI, H-1525 Budapest, POB 49, Hungary}
\begin{abstract}
We revisit the question of whether the $\mathrm{U}_A(1)$ symmetry is effectively restored in hot and dense medium.  In particular, by generalizing the Witten-Veneziano formula to finite temperature, we investigate whether the mass of $\eta^\prime$-meson will change in medium due to the restoration of chiral symmetry.
\end{abstract}


\maketitle

\section{Introduction}

The breaking of the  $\mathrm{U}_A(1)$ symmetry is an operator relation
that remains valid even when the spontaneously broken chiral symmetry is
restored. However, whether its effect on the $\eta'$ mass survives even when chiral
symmetry is restored is a phenomenological question that has caught the
interest of many
researchers \cite{Bernard:1987sx,Shuryak:1993ee,Kapusta:1995ww,Cohen:1996ng,Lee:1996zy,Evans:1996wf}.
The question has recently been revived as the RHIC data on two pion
Bose-Einstein correlation at $\sqrt{s}=200$ GeV Au+AU collision seems to
suggest the quenching of the $\eta'$ mass in
medium \cite{Vertesi:2009wf,Vertesi:2009wf-2,Csorgo:2009pa}.  Its partial quenching in
nuclear medium is also of great interest as such effects could be probed
in finer detail in nuclear target
experiments \cite{Jido:2011pq,Nagahiro:2011fi,Nanova:2011jk}.

The restoration  of $\mathrm{U}_A(1)$ symmetry  will depend on two
important ingredients;  its relation to  chiral symmetry breaking and
the effects from topological configurations. How the former and latter
contribute to the $n$-point correlation functions
have been clarified in references \cite{Cohen:1996ng} and \cite{Lee:1996zy,Evans:1996wf}.  Combining both effects, it was
shown\cite{Lee:1996zy} that in the chiral limit, with $N_f$ flavors, the
symmetry will effectively be restored in correlation functions composed
of up to $N_f-1$ points \cite{Birse:1996dx}.
This means, for example, that for $N_f=3$,  $\mathrm{U}_A(1)$ symmetry will be
effectively restored in the two-point functions when chiral symmetry is
restored.  Still, the argument is based on correlation functions and
does not explicitly relate the $\eta'$ mass to the other pseudo-scalar
masses.  To establish this relation, we revisit the Witten-Veneziano
(WV) formula \cite{Witten:1979vv,Veneziano:1979ec} for the $\eta'$ mass
in vacuum and generalize it to finite temperature.  Although the formula
is obtained in the large $N_c$ limit, we will obtain an explicit
relation that relates the $\eta'$ mass to condensates and two-point
correlation functions at finite temperature.  Therefore the symmetry
restoration pattern observed in the two-point function will be reflected
in the $\eta'$ mass.  For the physical case of $N_f=3$ with explicit
quark masses, this result implies that the mass of $\eta'$ will become
degenerate with the other pseudo-scalar mesons up to the quark masses
when chiral symmetry is restored.

The paper is organized as follows.  In section
\ref{sec:correlation}, we will revisit the
main results in ref.~\cite{{Lee:1996zy}}.  We will then review and
generalize the WV formula in section \ref{sec:wv}. We discuss the $\eta'$
mass in Sec.IV. The summary will be given in
the last section.

\section{Correlation functions and the $\mathrm{U}_A(1)$ symmetry}
\label{sec:correlation}  

Here, we start with a brief summary of the main result given in
ref.\cite{Lee:1996zy}.  The starting point is the Euclidean partition
function of QCD:
\begin{eqnarray}
Z[J] & = & \int D[A] e^{-S_{YM}} {\rm Det} [D \hspace*{-.2cm}\slash +m_q] \nonumber \\
& =& \sum_{\nu} Z[J]_\nu,
\end{eqnarray}
where $S_{YM}=(1/4)F^2$.  The second line writes the partition function
in terms of topological configurations with the topological charge
$\nu= (g^2/32) \int d^4x F \tilde{F}$.  The whole
integral in the first line is a positive definite
quantity \cite{Cohen:1996ng}, which we will denote as $d\mu$ for later convenience.
The topological configurations are always accompanied by $n_+ (n_-)$
number of right-handed (left-handed) fermion zero modes such that
$\nu=n_+-n_-$.  In such topologically non-trivial configurations, the
fermion determinant comes with special chirality such that partition
function with $\nu=1$ can be written more explicitly as follows:
\begin{eqnarray}
Z[J]_{\nu=1}  & = & \int D[A]_{\nu=1} e^{-S_{YM}} {\rm Det'} [D \hspace*{-.2cm}\slash +m_q] \nonumber \\
&  & \times {\rm det} \bigg( \int d^4x \bar{\psi_0}(x) m_q \psi_0(x) \bigg) ,
\label{part-top}
\end{eqnarray}
where the prime in the fermion determinant means that the chiral zero
modes $\psi_0$ have been explicitly taken out into the second
determinant.
Therefore, in the chiral limit $m_q=0$, the topological
configuration does not contribute to the partition function as
the fermion determinant gives zero.  However, these terms do contribute
in the correlations functions and select  out the $\eta'$ from the other
pseudo-scalars.  Higher topological configurations will contribute at higher point functions when there are sufficiently many external legs to saturate the zero modes.

To see this, consider a  two point function of a generic quark bilinear:
\begin{align}
\Pi_\Gamma(x) = & \langle \bar{q}(x) \Gamma q(x),  \bar{q}(0) \Gamma q(0) \rangle \nonumber  \\
= & \frac{1}{Z} \int d \mu \bigg[ - {\rm Tr} [S_A(x,0) \Gamma S_A(0,x)\Gamma] \nonumber \\ & + {\rm Tr} [S_A(x,x) \Gamma]  {\rm Tr} [S_A(0,0) \Gamma] +(zero~mode) \bigg],
\end{align}
the first, second and third term being the connected, disconnected terms
and  possible zero mode contribution respectively. $S_A$ is the quark
propagator in the presence of the gauge field.  When $\Gamma$ contains a
flavor matrix, the contributions from the disconnected diagrams are
identically zero.

The results in refs.\cite{Cohen:1996ng} and \cite{Lee:1996zy} can
be summarized as follows.  When one takes the difference between the two-point functions of chiral partners, the difference vanishes when chiral
symmetry is restored.  When the difference is taken between those composed of currents that are
related by a chiral transformation and an extra $\mathrm{U}_A(1)$
transformation, there will be an extra contribution from the zero modes.
As an example, the difference between a pseudo-scalar and $\eta'$ is
given as follows in SU(2):
\begin{align}
\Pi_\pi(x) - & \Pi_{\eta'}(x)= \frac{1}{Z} \int d \mu \bigg[   {\rm Tr} [S_A(x,x) \gamma_5]  {\rm Tr} [S_A(0,0) \gamma_5 ]\bigg] \nonumber \\
 & +\frac{1}{Z} \int d \mu_{\nu=\pm 1} \bigg[4 \bar{\psi_0}(x) \psi_0(x) \bar{\psi_0}(0) \psi_0(0) \bigg]. \label{pi-eta-diff}
\end{align}
In ref.\cite{Cohen:1996ng}, it was shown that the first term goes to
zero in the chiral limit when chiral symmetry is restored.   This result
is in fact independent of the number of flavors and also valid when the $\gamma_5$ inside the trace is replaced by other gamma matrices such as $\gamma_\mu$ or $\gamma_\mu \gamma_5$.

The zero-mode contributions appearing in the second term of
Eq.(\ref{pi-eta-diff}) come from the topological configuration in
Eq.(\ref{part-top}) and are responsible for the appearance of the
$\mathrm{U}_A(1)$ effect.  However, when $N_f > 2$, the second
determinant in Eq.(\ref{part-top}) will have 2$N_f$ zero mode lines and
hence the zero-mode contributions in Eq.(\ref{pi-eta-diff}) will be
proportional to $O(m_q^{N_f-2})$ and vanish in the chiral limit.
Therefore, when chiral symmetry is restored, the $\mathrm{U}_A(1)$
breaking effect will not appear in the two-point functions. However,
this does not necessarily mean that the $\eta'$ mass will become
degenerate with the other pseudo scalar mesons because the coupling of
the  currents to the $\eta'$ might just go to zero.   Therefore, let us
look at a relation that directly relates the mass of $\eta'$ to the chiral
order parameters.

\section{Witten-Veneziano Formula}
\label{sec:wv}

\subsection{WV formula at zero temperature}

As a first step of this study, we review the derivation of the WV mass
formula~\cite{Witten:1979vv}.  We start with the gluonic correlation
function defined in a pure glue theory:
\begin{equation}
	U(k)=i\int\ud^4x\,e^{ik\cdot x}\bra\mathcal{T}G\tilde{G}(x)\,G\tilde{G}(0)\ket.
	\label{eq1}
\end{equation}
One should note that,  in the large $N_c$ limit\cite{'tHooft:1973jz}, \eqn{eq1} scales as order
$N_c^2$.  There is also a well known low energy theorem for the
correlation function at zero external momentum $U(k=0)\neq 0$, whose value we
will come back in the next section.

However, when massless quarks are added to the theory, the low energy
theorem leads to the vanishing correlation function $U_{lq} (k=0)=0$, where the subscript
$lq$ means the presence of light quarks, through the anomaly relation that
relates the pseudo-scalar gluon current to
axial current
$\sum_q \partial_\mu \bar{q} \gamma_\mu \gamma_5 q = N_f \frac{\alpha_s }{4\pi}
G \tilde{G}$, where $\tilde{G}^a_{\mu \nu}=1/2 \epsilon_{\mu \nu \alpha
\beta}  G^a_{\alpha \beta}$.
This seems a little odd, because quark effects are
suppressed in large $N_c$, but for the low energy theorem, the leading
$N_c$ effect seems to be canceled by a subleading $N_c$ effect.  The
answer to this question led to the WV formula.

In terms of the physical states, the correlation functions looks as follows  when light quarks are added.
\begin{equation}
\begin{split}
	U_{lq}(k)&=-\sum_{n}\frac{|\bra0|G\tilde{G}|\,\text{$n^\textrm{th}$ glueball}\ket|^2}{k^2-M_n^2}\\
		  &\qquad-\sum_{n}\frac{|\bra0|G\tilde{G}|\,\text{$n^\textrm{th}$ meson}\ket|^2}{k^2-m_n^2}\\
		  &\equiv U_0(k)+U_1(k).
\end{split}
	\label{eq2}
\end{equation}
In the spectral form, \eqn{eq2}, the first term in the right hand side indicates contributions from glueballs,  while the second term shows those from the meson composed of light quarks.
One can show that the residue of the first term is of order $N_c^2$  whereas the quark effects are of order $N_c$\cite{Witten:1979kh}:
\begin{equation}
\begin{split}
	\bra0|G\tilde{G}|\,\text{$n^\textrm{th}$ glueball}\ket&=N_c\,a_n,\\
	\bra0|G\tilde{G}|\,\text{$n^\textrm{th}$ meson}\ket&=\sqrt{N_c}\,c_n.
\end{split}
	\label{eq3}
\end{equation}
Since all the meson masses should have a smooth large $N_c$ limit $O(1)$, the terms of quark effects, $U_1$, are suppressed in $1/N_c$ as expected.
The resolution to the seemingly inconsistent result is by noting the existence of  $\eta^\prime$-meson whose mass scales as order $1/N_c$.
One then recovers the
consistent low energy theorem  $U_{lq}(0)=U_0(0)+U_1(0)=0$, if the second term is saturated by the $\eta'$-meson and scales as $N_c^2$.   From this condition, one finds,
\begin{equation}
U(0)=	 U_0(0)=-\frac{|\bra0|G\tilde{G}|\eta^\prime\ket|^2}{m_{\eta^\prime}^2}=-\frac{N_c\,c_{\eta^\prime}^2}{m_{\eta^\prime}^2}.
	\label{eq4}
\end{equation}
By using the U$(1)_A$ anomaly,
\begin{equation}
\begin{split}
	 \bra0|G\tilde{G}|\eta^\prime\ket&=\frac{4\pi}{\alpha}\frac{1}{N_f}\bra0|\partial_\mu J_5^\mu|\eta^\prime\ket\\
					 &=\frac{4 \pi}{\alpha}\frac{1}{N_f}\sqrt{N_f}\,m_{\eta^\prime}^2f_\pi,
\end{split}
	\label{eq5}
\end{equation}
\eqn{eq4} becomes as follows:
\begin{equation}
	 U_0(0)=\frac{1}{N_f}m_{\eta^\prime}^2f_\pi^2\left(\frac{4\pi}{\alpha}\right)^2,
	\label{eq6}
\end{equation}
where $N_f$ is the number of light flavors. In \eqn{eq5}, we made use of $f_{\eta^\prime}=f_\pi$ to lowest order in $N_c$.  \eqn{eq6} is the celebrated WV formula.

\subsection{WV formula at finite temperature}

\begin{figure}[bh]
 \includegraphics[width=7cm,height=4.5cm]{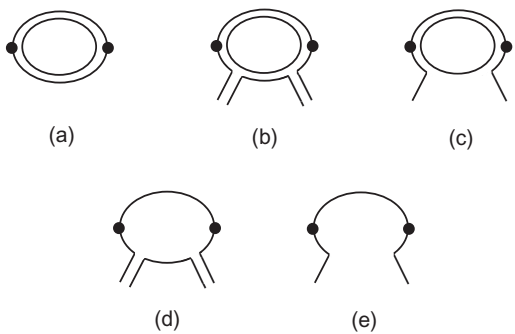}
 \caption{Two sets of diagrams}
	\label{fig1}
\end{figure}

Consider the correction to \eqn{eq6} at finite temperature. As mentioned
before, the correlation function in \eqn{eq1} is order $N_c^2$, as can
be seen by the two loops representing two gluon lines in \fig{fig1}-(a).
At finite temperature, the thermal correction could come from the
thermal gluon or quark interactions. \fig{fig1}-(b,c) show the thermal corrections to $U_0(k)$ while \fig{fig1}-(d,e) show those to $U_1(k)$.

The dominant thermal gluonic contribution to $U_0(k)$ comes from  \fig{fig1}-(b) and scales as $N_c^2$ as in the vacuum scaling.  The scaling
comes as follows,
\begin{equation}
	(N_c) \times \left(\frac{1}{\sqrt{N_c}} \right)^2 (N_c)^2 =O(N_c^2).
	\label{eq7}
\end{equation}
which comes from the internal loop, two coupling and the number of
external thermal gluons respectively.
The contributions where the gluons couple directly to the currents scale the same as in \fig{fig1}-(b).
On the other hand, the contributions from thermal quarks to $U_0(k)$ scales as follows,
\begin{equation}
	(N_c) \times \left(\frac{1}{\sqrt{N_c}} \right)^2 (N_c) =O(N_c),
	\label{eq8}
\end{equation}
where the factors are the same as before except for the last factor,
which comes from the number of external quark lines, as can be seen from
\fig{fig1}-(c).
Therefore, in the large $N_c$ limit, the thermal gluonic effect scales
the same as in the leading vacuum scaling and will contribute to
modifying $U_0(k)$.

As for the modification in the quark loops $U_1(k)$, the thermal gluonic effects are shown in \fig{fig1}-(d), and the thermal quark effects in \fig{fig1}-(e).  Both scale as $O(N_c)$, and can thus be neglected in the leading order correction.

If the system is in the confined state, the hadronic
side will be saturated by color singlet glueball, meson and nucleons.
Here, the dominant contribution comes from the nucleons.
One can show that the contribution from the nucleon to all figures in \fig{fig1} scale as $O(N_c)$ because the nucleon contains $N_c$ quarks.  On the other hand, the contributions from meson or glueballs are suppressed in $1/N_C$ as the number of constituents are finite.   Hence, hadron effects can be neglected until near the phase transition point where the density of states increases, after which one can use the quark and gluon degrees of freedom.

Therefore, same arguments hold as in the vacuum.
Namely, the addition of quarks somehow has to cancel the leading $N_c$
behavior at $k=0$.  This cancelation can not be done by collective
states, as quark collective states are also suppressed in large $N_c$
limit, and hence has to come from a modified $\eta^\prime$ contribution.
All in all, a similar equation to \eqn{eq4} will hold at
finite temperature, with $|\bra0|G\tilde{G}|\eta^\prime\ket|$ now defined at finite
temperature at $\eta^\prime$ momentum zero.
Moreover, it should be noted that the $\eta^\prime$ mass we are
discussing now could be different from that of the pole mass as we are
discussing the scalar part of the mass, which survives at $k^\mu
\rightarrow 0$.  A simplified example would be to assume that the small
energy and momentum self energy has the following form, with
$a(T)$ and $b(T)$ being the small corrections,
\begin{equation}
	\Sigma_{\eta^\prime}=a(T) k_0^2+b(T) \vec{k}^2 + m^2(T).
	\label{eq9}
\end{equation}
The pole mass at $\vec{k} \rightarrow 0$ would be $\sqrt{m^2+m^2(T)}/\sqrt{1-a(T)}$.  But the mass we are talking about is $\sqrt{m^2+m^2(T)}$.

Nevertheless, $U_0$ has a nontrivial correction at finite temperature as
we will see in the following sections.

\section{$\eta'$ mass at finite temperature}

\subsection{Low energy theorem}
Now, $U_0(0)$ can be obtained from the low energy theorem.  Here we use the derivation using the heavy quark expansion~\cite{Lee:2000cdb}.
For technical reasons, we start from a slightly different definition of the correlation function.
\begin{equation}
	P(k)=i \int \ud^4x \,e^{ik\cdot x}\left\bra \mathcal{T}\left[\frac{ 3 \alpha}{4 \pi} G  \tilde{G}(x) , \frac{ 3 \alpha}{4 \pi} G \tilde{G}(0)\right] \right\ket
	\label{eq10}
\end{equation}

It can be shown~\cite{Lee:2000cdb} that
\begin{equation}
	P(k=0)=-\frac{2}{32 \pi^2} \frac{\ud}{\ud(-1/4g_0^2)} \left\bra \frac{\alpha}{\pi} G^2 \right\ket.
	\label{eq11}
\end{equation}

Now, any matrix element with canonical dimension $d$ should be proportional to the $d^{\,\mathrm{th}}$ power of the scale
\begin{equation}
	\Lambda=M_0 \exp \left( -\frac{8 \pi^2}{bg_0^2} \right)
	\label{eq12}
\end{equation}
with $b=11-\frac{2}{3}N_f$.
Hence, the gluon condensate at finite temperature and density should be of the following form.
\begin{equation}
	\left\bra \frac{\alpha}{\pi} G^2 \right\ket_{T,\mu} = \Lambda^d f \left(\frac{T}{\Lambda},\frac{\mu}{\Lambda} \right),
	\label{eq13}
\end{equation}
where $d=4$ and  $f$ is a generic function specifying the temperature and density dependence of the gluon condensate.
Then \eqn{eq11} becomes,
\begin{eqnarray}
	P(k=0)=-\frac{2}{b} \left(d - T \frac{\partial }{\partial T} -\mu \frac{\partial }{\partial \mu}\right) \left\bra \frac{\alpha}{\pi} G^2 \right\ket_{T,\mu}.
	\label{eq14}
\end{eqnarray}

Now, combining \eqn{eq4} and \eqn{eq14}, one finds,
\begin{equation}
\bigg( \frac{3 \alpha}{4 \pi} \bigg)^2
\frac{ | \langle 0 | G  \tilde{G} |\eta^\prime \rangle |^2}{ m_{\eta^\prime}^2}
=\frac{2}{b}  \left(d - T \frac{\partial }{\partial T} -\mu \frac{\partial }{\partial \mu} \right) \left\bra \frac{\alpha}{\pi} G^2 \right\ket_{T,\mu}.
	\label{eq15}
\end{equation}

But now since we can make the identification of the left and right hand
side only in the large $N_c$ limit, the right hand side should be
calculated in the quenched approximation.  This means that one should just read off the temperature dependence of the gluon condensate from the lattice calculation for pure gauge theory, and also
take $b=11$.
Thus, the $\eta'$ mass is given by
\begin{equation}
 m_{\eta^\prime}^2 = \left( \frac{3\alpha}{4\pi} \right)^2
  \frac{|\langle 0|
  G\tilde{G}|\eta^\prime \rangle |^2}{\frac{2}{b}\left(
						  d-T\frac{\partial}{\partial
						  T} \right)
  \left\langle \frac{\alpha}{\pi}G^2 \right\rangle_{T,\text{pure
  gauge}}}.
  \label{eq16}
\end{equation}

\subsection{Gluonic part}
In order to evaluate the in-medium $\eta^\prime$ mass from \eqn{eq16},
all we need are the temperature-dependence of the gluon condensate and
the coupling of $G\tilde{G}$ to $\eta'$.
First let us consider the denominator of \eqn{eq16}.
It has been known for a long time, that the gluon condensate has
contribution from the perturbative and non-perturbative contribution.
Moreover, it was also known that at the critical temperature, the
non-perturbative contribution changes abruptly, but does not vanish
completely, and retains more than half of its non-perturbative
value~\cite{Lee:1989qj,Adami:1990sv,Brown:2006bp}.


The effect of subtracting out the second term in the denominator of \eqn{eq16} is to get rid of the perturbative
correction, or the seemingly scale breaking effect that is not related
to scale breaking but due to the introduction of an external scale
parameter $T$.  The leading perturbative correction to the gluon
condensate is proportional to
$g^4(T)T^4$~\cite{Kapusta:1979fh,Boyd:1996bx}.  Therefore, assuming that
the temperature dependence is of the following form,
\begin{equation}
	\left\bra \frac{\alpha}{\pi} G^2 \right\ket_T=G_0(T) +ag^4T^4,
	\label{eq17}
\end{equation}
we find,
\begin{equation}
	\left(d - T \frac{\partial }{\partial T}\right) \left\bra \frac{\alpha}{\pi} G^2 \right\ket_T =\left(d - T \frac{\partial }{\partial T}\right) G_0(T),
	\label{eq18}
\end{equation}
if the temperature dependence  of g is neglected\footnote{At
low temperature this may not be true. But since $dg(T)/dT < 0$, the
conclusion will not change if we consider this seriously.}.
The only temperature dependence that survives is $G_0(T)$, whose scale
dependence is coming from dimensional transmutation and not from the
external temperature only.
It is the non-perturbative part that dominates the behavior of the right
hand side of \eqn{eq16}.
\begin{figure}[t]
 \includegraphics[width=7cm,height=4.5cm]{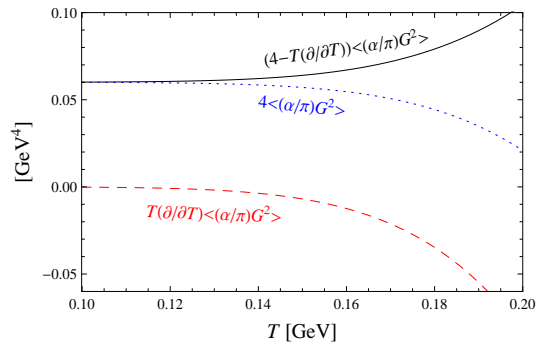}
\caption{$T$-dependence of gluon condensate and its derivative fitted to Wuppertal-Budapest lattice data in full QCD.}
	\label{fig2}
\end{figure}


In \fig{fig2}, we show  the model fit~\cite{Morita:2010pd} to the Wuppertal-Budapest's full QCD data~\cite{Borsanyi:2010cj} together with the modified $T$-dependent gluon operator as it appears in \eqn{eq16}.   In the quenched calculation, the only difference is that the change is taking place more abruptly near the new critical temperature\cite{Morita:2007hv}.   One can see that the change in the denominator of \eqn{eq16} will tend to reduce the $\eta'$ mass near the critical temperature.

\subsection{coupling to $\eta'$}

The final step in obtaining the mass of $\eta^\prime$ using \eqn{eq16}, when chiral symmetry is restored, is estimating the change of the  coupling $\langle 0 | G  \tilde{G} |\eta^\prime \rangle$.  For that purpose, let us consider $U_{lq}(k)$ in \eqn{eq1} in the full theory, but rewrite it in terms of the quark axial current using the anomaly relation.
\begin{align}
U(k)  =& i\int\ud^4x\,e^{ik\cdot x}\bra\mathcal{T}G\tilde{G}(x)\,G\tilde{G}(0)\ket \nonumber \\
 = & k^\mu k^\nu i\int\ud^4x\,e^{ik\cdot x}
\bigg( \frac{4 \pi }{ \alpha N_f} \bigg)^2 \bigg[\bra\mathcal{T} \bar{q} i \gamma_\mu \gamma_5 q(x) \,
\bar{q} i \gamma_\nu \gamma_5 q(0) \ket
 \nonumber \\
& -
\bra\mathcal{T} \bar{q}  \gamma_\mu q(x) \,
\bar{q}  \gamma_\nu  q(0) \ket \bigg], \label{coupling}
\end{align}
where we have subtracted out the contribution from the conserved vector current.  Using the previous terminology, when chiral symmetry is restored, the connected piece will cancel, as they are the same as the difference between flavored chiral partners, and only the disconnected pieces will remain.  Assuming that the spectral sum starts from the $\eta^\prime$, we find \eqn{coupling} can then be written as follows,
\begin{widetext}
\begin{align}
U(k) =& - \frac{|\bra 0 |G\tilde{G}| \eta^\prime \ket |^2}{k^2-m_{\eta^\prime}^2}   - \cdots   \nonumber \\
\rightarrow & k^\mu k^\nu \int \bigg( \frac{4 \pi }{ \alpha N_f} \bigg)^2 \bigg[   {\rm Tr} [S_A(x,x) i \gamma_\mu \gamma_5]
 {\rm Tr} [S_A(0,0)i \gamma_\nu \gamma_5 ]
-  {\rm Tr} [S_A(x,x)  \gamma_\mu ]  {\rm Tr} [S_A(0,0) \gamma_\nu  ]  \bigg].
\label{sumrule}
\end{align}
\end{widetext}
However, the disconnected pieces are all of the same order in $m_q$ when chiral symmetry is restored.
\begin{eqnarray}
{\rm Tr} [S_A(x,x)] \sim {\rm Tr} [S_A(x,x) \Gamma ] \sim O(m_q),
\end{eqnarray}
where $\Gamma$ is a Hermitian gamma matrix\cite{Cohen:1996ng}.
Since \eqn{sumrule} is valid for any $k$, we find
that
\begin{eqnarray}
\bra 0|G\tilde{G} | \eta^\prime \ket \sim O(m_q),
\end{eqnarray}
when chiral symmetry is restored.  Therefore, going back to
\eqn{eq16} and making use of the previous discussions,
we find that when chiral symmetry is restored,
\begin{eqnarray}
m_{\eta^\prime}^2
\stackrel{\bra \bar{q}q \ket \rightarrow 0}{\longrightarrow}  0 , \label{final}
\end{eqnarray}
in the chiral limit.
One concludes that in the large $N_c$ limit of QCD, $\eta^\prime$
mass will become degenerate with the other goldstone bosons.

\section{Conclusions}

It should be noted that the $\eta^\prime$ mass that is being quenched is
the part of the mass that comes from the breaking of the $\mathrm{U}_A(1)$ symmetry.
Going back to \eqn{eq6} and substituting the vacuum value of \eqn{eq14}
one finds,
\begin{eqnarray}
m_{\eta^\prime} =\sqrt{\frac{8}{33}} \frac{1}{f_\pi} \bra \frac{\alpha}{\pi}G^2 \ket^{1/2} \approx 464 ~~{\rm MeV},
\end{eqnarray}
where we have used $f_\pi=130\,\mathrm{MeV}$ and $ \bra \frac{\alpha}{\pi}G^2 \ket= (0.35 {\rm GeV})^4$.  This is smaller than the vacuum value of the $\eta^\prime$ mass as expected.   Assuming that the pseudo scalar mesons do not change their mass towards the phase transition point, it is this extra $\mathrm{U}_A(1)$ mass of $\eta^\prime$  that is going to be quenched in the chiral symmetry restored phase.

Few remarks are in order.  First, in the quenched approximation, the changes of order parameters take place only near the phase transition point.  This suggests that the effect of quenching might only be visible when the hadronization temperature is close to the phase transition point as in the case of  RHIC or LHC  energies for example.
Second, it is hard to make a quantitative estimate on how much of this mass is quenched in the nuclear medium, as the effects of density are  subleading in the large $N_c$ limit.  However, assuming \eqn{eq16} is an exact relation, we can use \eqn{sumrule} to approximate  $\langle 0 | G  \tilde{G} |\eta^\prime \rangle \propto {\rm Tr}[S_A(x,x)]\propto \langle \bar{q}q \rangle $ and then use it in \eqn{eq16} to deduce $m_{\eta^\prime} \propto \langle \bar{q}q \rangle$, assuming that the change in the gluon condensate is small in nuclear medium.  Therefore, if the chiral order parameter reduces by 20\% in nuclear medium the $\mathrm{U}_A(1)$ breaking part of the $\eta^\prime$ mass will also reduce by the same fraction.

\section*{ACKNOWLEDGEMENTS}

This work was supported by Korea national research foundation under grant number KRF-2011-0030621.
YK was supported (in part) by the Yonsei University Research Fund of 2010 and by Korea national research foundation under grant number KRF-2011-0015467.
KM is supported by Yukawa International Program for Quark-Hadron
Sciences at YITP, Kyoto university. GyW was supported by the Hungarian OTKA T71989 and T101438.


\end{document}